\documentclass[12pt]{article}
\begin{document}

\title{\bf Energy-Momentum Distribution: Some Examples}
\author{M. Sharif \thanks{e-mail: msharif@math.pu.edu.pk} and M. Azam
\\ Department of Mathematics, University of the Punjab,\\ Quaid-e-Azam
Campus Lahore-54590, PAKISTAN.}

\date{}

\maketitle

\begin{abstract}
In this paper, we elaborate the problem of energy-momentum in
General Relativity with the help of some well-known solutions. In
this connection, we use the prescriptions of Einstein,
Landau-Lifshitz, Papapetrou and M\"{o}ller to compute the
energy-momentum densities for four exact solutions of the Einstein
field equations. We take the gravitational waves, special class of
Ferrari-Ibanez degenerate solution, Senovilla-Vera dust solution and
Wainwright-Marshman solution. It turns out that these prescriptions
do provide consistent results for special class of Ferrari-Ibanez
degenerate solution and Wainwright-Marshman solution but
inconsistent results for gravitational waves and Senovilla-Vera dust
solution.
\end{abstract}


\date{}

\section{Introduction}

In the theory of General Relativity (GR), the energy-momentum
conservation laws are given by
\begin{equation}
T^b_{a;b}=0,\quad(a,b=0,1,2,3),
\end{equation}
where $T^{b}_a$ denotes the energy-momentum tensor. In order to
change the covariant divergence into an ordinary divergence so that
global energy-momentum conservation, including the contribution from
gravity, can be expressed in the usual manner as in
electromagnetism, Einstein formulated [1] the conservation law in
the following form
\begin{equation}
\frac{\partial}{\partial x^b}(\sqrt{-g}(T^b_a+t^b_a))=0.
\end{equation}
Here $t^b_a$ is not a tensor quantity and is called the
gravitational field pseudo-tensor. Schrodinger showed that the
pseudo-tensor can be made vanish outside the Schwarzschild radius
using a suitable choice of coordinates. There have been many
attempts in order to find a more suitable quantity for describing
the distribution of energy and momentum due to matter,
non-gravitational and gravitational fields. The proposed quantities
which actually fulfill the conservation law of matter plus
gravitational parts are called gravitational field pseudo-tensors.
The choice of the gravitational field pseudo-tensor is not unique.
Because of this, quite a few definitions of these pseudo-tensors
have been proposed.

Misner at el. [2] showed that the energy can only be localized in
spherical systems. But later on, Cooperstock and Sarracino [3]
proved that if energy is localizable for spherical systems, then it
can be localized in any system. Einstein was the first to construct
a locally conserved energy-momentum complex [4]. After this attempt,
many physicists including Tolman [5], Landau-Lifshitz [6],
Papapetrou [7], Bergmann [8] and Weinberg [9] introduced different
definitions for the energy-momentum complex. These definitions can
only give meaningful results if the calculations are performed in
Cartesian coordinates. In 1990, Bondi [10] argued that a
non-localizable form of energy is not allowed in GR. After this, the
idea of quasi-local energy was introduced by Penrose and other
researchers [11-13]. In this method, one can use any coordinate
system while finding the quasi-local masses to obtain the
energy-momentum of a curved spacetime. Bergqvist [14] considered
seven different definitions of quasi-local mass and showed that no
two of these definitions give the same result. Chang at el. [15]
showed that every energy-momentum complex can be associated with a
particular Hamiltonian boundary term and hence the energy-momentum
complexes may also be considered as quasi-local.

M\"{o}ller [16,17] proposed an expression which is the best to make
calculations in any coordinate system. He claimed that his
expression would give the same results for the total energy and
momentum as the Einstein's energy-momentum complex for a closed
system. Lessner [18] gave his opinion that M\"{o}ller's definition
is a powerful concept of energy and momentum in GR. However,
M\"{o}ller's prescription was also criticized by some people
[10,19-21]. Komar's complex [21], though not restricted to the use
of Cartesian coordinates, is not applicable to non-static
spacetimes. Thus each of these energy-momentum complex has its own
drawbacks. As a result, these ideas of the energy-momentum complexes
could not lead to some unique definition of energy in GR.

Scheidegger [22] raised doubts whether gravitational radiation has
well-defined existence. Rosen [23] investigated whether or not
cylindrical gravitational waves have energy and momentum. He used
the energy-momentum pseudo-tensors of Einstein and Landau-Lifshitz
and carried out calculations in cylindrical coordinates. He found
that the energy-momentum densities vanish. The results obtained fit
in the conjecture of Scheidegger that a physical system cannot
radiate gravitational energy. Two years later, Rosen [24] realized
the mistake and carried out the calculations in Cartesian
coordinates. He found that energy-momentum densities are
non-vanishing and reasonable. After that Rosen and Virbhadra [25]
calculated the energy-momentum densities of gravitational waves in
Einstein complex and found to be finite and reasonable results.
Numerous attempts have been made to resolve the problem of
energy-momentum localization but still remains un-resolved. This
problem first appeared in electromagnetism which turns out to be a
serious matter in GR due to the non-tensorial quantities.

Virbhadra et al. [25-31] explored several spacetimes for which
different energy-momentum complexes show a high degree of
consistency in giving the same and acceptable energy-momentum
distribution. Aguirregabira et al. [32] showed that five different
energy-momentum complexes gave the same result for any Kerr-Schild
class (including the Schwarzchild, Reissner-Nordstr\"{o}m, Kerr and
Vaidya metrics). Xulu [33-35] extended this investigation and found
same energy distribution in the Melvin magnetic and Bianchi type I
universe. Chamorro and Virbhadra [36] and Xulu [37] studied the
energy distribution of charged black holes with a dilaton field.

This paper explores some more examples to investigate this problem.
The paper is organized as follows. In section 2, we shall briefly
mention different prescriptions to evaluate energy-momentum
distribution. Sections 3-6 are devoted for the evaluation of
energy-momentum densities for the four particular spacetimes. The
last section contains summary of the results obtained.

\section{Energy-Momentum Complexes}

We shall use four different prescriptions, i.e., Einstein,
Landau-Lifshitz, Papapetrou and M\"{o}ller, to evaluate the
energy-momentum density components of different spacetimes. For the
sake of completeness, we briefly give these formulae as details are
available elsewhere [38-41].

The energy-momentum complex of Einstein [42] is given by
\begin{equation}
\Theta^{b}_{a}=\frac{1}{16\pi}H^{bc}_{a,c},\quad
(a,b,...=0,1,2,3),
\end{equation}
where
\begin{equation}
H^{bc}_a=\frac{g_{ad}}{\sqrt{-g}}[-g(g^{bd}g^{ce}-g^{cd}g^{be})]_{,e}.
\end{equation}
It is to be noted that $H^{bc}_a$ is anti-symmetric in indices $b$
and $c$. $\Theta^{0}_{0}$ is the energy density,
$\Theta^{i}_{0}~(i=1,2,3)$ are the components of momentum density
and $\Theta^{0}_{i}$ are the energy current density components.

The prescription of Landau-Lifshitz [6] is defined
as
\begin{equation} L^{ab}= \frac{1}{16 \pi}\ell^{acbd}_{,cd},
\end{equation}
where
\begin{equation}
\ell^{acbd}= -g(g^{ab}g^{cd}-g^{ad}g^{cb}).
\end{equation}
$L^{00}$ represents the energy density of the whole system including
gravitation and $L^{oi}$ represent the components of the total
momentum density. $\ell^{abcd}$ has symmetries of the Riemann
curvature tensor. It is clear from Eq.(7) that $L^{ab}$ is symmetric
with respect to its indices.

The symmetric energy-momentum complex of Papapetrou [7] is given as
\begin{equation}
\Omega^{ab}=\frac{1}{16\pi}N^{abcd}_{,cd},
\end{equation}
where
\begin{equation}
N^{abcd}=\sqrt{-g}(g^{ab}\eta^{cd}-g^{ac}\eta^{bd}
+g^{cd}\eta^{ab}-g^{bd}\eta^{ac})
\end{equation}
and $\eta^{ab}$ is the Minkowski spacetime. The quantities
$N^{abcd}$ are symmetric in its first two indices $a$ and $b$. The
locally conserved quantities $\Omega^{ab}$ contain contribution from
the matter, non-gravitational and gravitational field. The quantity
$\Omega^{00}$ represents energy density and $\Omega^{0i}$ are the
momentum density components.

The coordinate independent prescription by M\"{o}ller [17] is
defined as
\begin{equation}
M^{b}_{a}=\frac{1}{8\pi}K^{bc}_{a,c},
\end{equation}
where
\begin{eqnarray}
K^{bc}_{a}=\sqrt{-g}(g_{ad,e}-g_{ae,d})g^{be}g^{cd}.
\end{eqnarray}
Here $K^{bc}_{a}$ is antisymmetric, $M^0_0$ is the energy density,
$M^i_0$ are the momentum density components and $M^0_i$ are the
energy current density components. We shall apply these
prescriptions to particular examples.

\section{Gravitational Waves}

The general line element of gravitational waves [46] is given by
\begin{equation}
ds^2=e^{-M}(dt^2-dx^2)-e^{-U}(e^{-V}dy^2+e^{V}dz^2),
\end{equation}
where $U,~V$ and $M$ are functions of $t$ and $x$ only. In the case
of a stiff perfect fluid, the Einstein field equations imply that
$U$ satisfies the wave equation
\begin{equation}
U_{tt}-U_{xx}=0
\end{equation}
and $U,~V$ satisfy the linear equation
\begin{equation}
V_{tt}-U_tV_t-V_{xx}+U_xV_x=0.
\end{equation}
It may be noted that the solution describing the closed FRW stiff
fluid model can be given by
\begin{eqnarray*}
e^{-U}&=&\sin {2t} \sin {2x} ,\quad V=\ln {\tan x},\\
M&=&-\ln {\sin {2t}}- \ln {\gamma},\quad \sigma=\sqrt{3}\ln\tan t,
\end{eqnarray*}
where $ 0<t<\frac{\pi}{2},~ 0<x<\frac{\pi}{2}$ and $\gamma$ is
constant. A stiff perfect fluid can be associated with a potential
$\sigma(t,x)$ such that the density and 4-velocity of the fluid are
given by
\begin{eqnarray*}
16\pi\rho=e^M(\sigma^2_t-\sigma^2_x), \quad
u_a=\frac{\sigma_a}{(\sigma_b\sigma^b)^{\frac{1}{2}}}
\end{eqnarray*}
and the fluid potential $\sigma$ satisfies
\begin{equation}
\sigma_{tt}-U_t\sigma_t-\sigma_{xx}+U_x\sigma_x=0.
\end{equation}
A gravitational wave with toroidal wavefront can be obtained by
taking [47]
\begin{eqnarray*}
U=-\ln t-\ln \rho,\quad V=\ln t-\ln \rho+\tilde{V}(t,\rho),
\end{eqnarray*}
where $\tilde{V}$ has the form
\begin{eqnarray*}
\tilde{V}(t,\rho)&=&\int_\frac{1}{2}^\infty \phi(k)(t\rho)^k H_k
(\frac{t^2+\rho^2-a^2}{2t\rho})dk
\end{eqnarray*}
with an arbitrary funcation $\phi(k)$ and
$$
M=\frac{1}{2k}a_k(t^2-\rho^2)(t\rho)^{k-1}H_{k-1}
-\frac{1}{2}(t\rho)^{2k}a_k^2
[k^2H_k^2-\frac{(t^2-\rho^2)^2}{4t^2\rho^2}H_{k-1}^2],
$$
where the dimension of $a_k$ is $L^{-2k}$. Now we calculate
energy-momentum distribution by using the four prescriptions.

For the Einstein prescription, we need the following non-zero
components of $H^{bc}_a$
\begin{eqnarray}
H^{01}_0&=&-H^{10}_0=2U'e^{-(2M+3U)},\\
H^{01}_1&=&2\dot{U}e^{-U},
\end{eqnarray}
where dot and prime mean differentiating w.r.t. $t$ and $x$
respectively. Using Eqs.(15)-(16) in Eq.(3), we obtain the following
components of energy, momentum and energy current densities
\begin{eqnarray}
\Theta^0_0&=&\frac{e^{-(2M+3U)}}{8 \pi}[U''-2U'M'-3U'^2],\\
\Theta^1_0&=&\frac{e^{-(2M+3U)}}{8
\pi}[U'(2\dot{M}+3\dot{U})-\dot{U'}],\\
\Theta^0_1&=&\frac{e^{-U}}{8\pi}(\dot{U'}-\dot{U}U'),\\
\Theta^0_2&=&0=\Theta^2_0=\Theta^0_3=\Theta^3_0.
\end{eqnarray}

The non-zero components of $\ell^{abcd}$ are used in the
Landau-Lifshitz complex
\begin{eqnarray}
\ell^{0011}&=&-e^{-2U},\\
\ell^{0101}&=&e^{-2U}.
\end{eqnarray}
Substituting these values in Eq.(5), it follows the components of
energy and momentum (energy current) densities in Landau-Lifshitz
prescription
\begin{eqnarray}
L^{00}&=&\frac{e^{-2U}}{8\pi}(U''-2U'^2),\\
L^{10}&=&L^{01}=\frac{e^{-2U}}{8\pi}(2U'\dot{U}-\dot{U'}),\\
L^{20}&=&L^{02}=0=L^{30}=L^{03}.
\end{eqnarray}

For Papapetrou prescription, the non-zero components of $N^{abcd}$
are the following
\begin{eqnarray}
N^{0011}&=&-2e^{-U},\\
N^{1001}&=&N^{1010}=e^{-U}.
\end{eqnarray}
When we make use of these values in Eq.(7), it yields the following
components of energy and momentum (energy current) densities
\begin{eqnarray}
\Omega^{00}&=&\frac{e^{-U}}{8\pi}(U''-U'^2),\\
\Omega^{10}&=&\Omega^{01}=\frac{e^{-U}}{8\pi}(\dot{U}U'-\dot{U'}),\\
\Omega^{20}&=&\Omega^{02}=0=\Omega^{30}=\Omega^{03}.
\end{eqnarray}

The following non-zero components of $K^{bc}_a$ are required in
M\"{o}ller prescription
\begin{equation}
K^{01}_0=-K^{10}_0=-M'e^{-U}.
\end{equation}
Consequently, the components of energy, momentum and energy current
densities become
\begin{eqnarray}
M^0_0&=&\frac{e^{-U}}{8\pi}(M'U'-M'')\\
M^1_0&=&\frac{e^{-U}}{8\pi}(\dot{M'}-\dot{U}M'),\\
M^0_1&=&0=M^0_2=M^2_0=M^3_0=M^0_3.
\end{eqnarray}
We have obtained energy-momentum distribution for a general line
element of the gravitational waves. The energy-momentum distribution
for the colliding and toroidal gravitational waves can be found by
substituting the corresponding values of $U,~V,~M$.

\section{Special Class of Ferrari-Ibanez  Degenerate Solution}

The special class of Ferrari-Ibanez degenerate solution [48] is
given by
\begin{equation}
ds^2=(1+\sigma\sin t)^2(dt^2-dz^2)-\frac{(1-\sigma\sin
t)}{(1+\sigma\sin t)}dx^2-\cos^2z(1+\sigma\sin t)^2dy^2,
\end{equation}
where $\sigma=\pm1$ is an arbitrary constant and $t,z$ are timelike
and spacelike coordinates respectively. If we take the coordinate
transformation
\begin{eqnarray*}
r=1+sint, \quad \tau=\sqrt{2}x,\quad\theta=\frac{\pi}{2}-z,\quad
\phi=\sqrt{2}y,\quad m=1,
\end{eqnarray*}
then the line element (38) reduces to the Schwarzschild metric. It
is to be noted that there is a curvature singularity for $\sigma=-1$
and $t=\frac{\pi}{2}$. However, the spacetime appears to be regular
for $\sigma=1$ and $0\leq t\leq\frac{\pi}{2}$.

The required components of $H^{bc}_a$ for Einstein complex are
\begin{eqnarray}
H^{03}_0&=&-H^{30}_0=2\sin z(1-\sigma^2\sin^2 t)^\frac{1}{2},\\
H^{03}_3&=&\frac{2\cos z\sin t\cos t}{(1-\sigma^2\sin^2
t)^\frac{1}{2}}.
\end{eqnarray}
Thus the components of energy, momentum and energy current densities
with $\sigma=\pm1$ become
\begin{eqnarray}
\Theta^0_0&=&\frac{1}{8 \pi}(\cos t\cos z),\\
\Theta^3_0&=&\frac{1}{8\pi}(\sin t\sin z)=-\Theta^0_3,\\
\Theta^0_1&=&\Theta^1_0=0=\Theta^0_2=\Theta^2_0.
\end{eqnarray}

For Landau-Lifshitz prescription, the non-zero components of
$\ell^{abcd}$ are as follows:
\begin{eqnarray}
\ell^{0033}&=&-cos^2z(1-\sigma^2\sin^2 t),\\
\ell^{0303}&=&cos^2z(1-\sigma^2\sin^2 t).
\end{eqnarray}
The components of energy and momentum (energy current) densities are
\begin{eqnarray}
L^{00}&=&\frac{1}{8\pi}(\cos t\cos z),\\
L^{30}&=&L^{03}=\frac{1}{16\pi}(\sin 2 t\sin 2 z),\\
L^{10}&=&L^{01}=0=L^{20}=L^{02}.
\end{eqnarray}

The energy-momentum densities in Papapetrou complex can be found by
using the components of $N^{abcd}$ as
\begin{eqnarray}
N^{0033}&=&-2\cos z(1-\sigma^2\sin^2t)^\frac{1}{2},\\
N^{3003}&=&N^{3030}=2\cos z(1-\sigma^2\sin t)^\frac{1}{2}.
\end{eqnarray}
As a result, the components of energy and momentum (energy current)
densities turn out to be
\begin{eqnarray}
\Omega^{00}&=&\frac{1}{8\pi}(\cos t\cos z),\\
\Omega^{30}&=&\Omega^{03}=\frac{1}{8\pi}(\sin t\sin z),\\
\Omega^{10}&=&\Omega^{01}=0=\Omega^{20}=\Omega^{02}.
\end{eqnarray}
We see that energy density becomes the same in three prescriptions.

For M\"{o}ller prescription, the required components of $K^{bc}_a$
are
\begin{equation}
K^{03}_3=-2\sigma\cos t\sin z\frac{(1-\sigma\sin
t)^\frac{1}{2}}{(1+\sigma\sin t)^\frac{1}{2}}
\end{equation}
and the components of energy, momentum and energy current densities
are
\begin{eqnarray}
M^0_0&=&0,\\
M^0_3&=&-\frac{1}{4\pi}\cos t\sin z\frac{(1-\sin
t)^\frac{1}{2}}{(1+\sin t)^\frac{1}{2}},\\
M^0_1&=&0=M^1_0=M^0_2=M^2_0=M^3_0.
\end{eqnarray}
This shows that energy and momentum become constant.

\section{Senovilla-Vera Dust Solution}

The Senovilla-Vera dust solution [49] in the fluid co-moving
coordinates is given by
\begin{equation}
ds^2=dt^2-t^2dx^2-Y^2dy^2-(te^{-x})^{1-k}dz^2,
\end{equation}
where
\begin{eqnarray*}
Y&=&c_{-}(te^{-x})^{b_{-}}+m^2(te^x)^b+c_{+}(te^{-x})^{b_{+}},\\
b&=&\frac{1}{2}(1+k), \quad b_{\pm}=\frac{1}{2}(1\pm\sqrt{2-k^2}),
\quad -1<k<1
\end{eqnarray*}
and $c_{-},~c_{+}$ are constants.

Einstein complex gives the components of $H^{bc}_a$ as
\begin{eqnarray}
H^{01}_0&=&(te^{-x})^{\frac{1-k}{2}}[2\{c_{-}b_{-}t^{b_{-}-1}e^{-xb_{-}}
-m^2bt^{b-1}e^{xb}\nonumber\\
&+&c_{+}b_{+}t^{b_{+}-1}e^{-xb_{+}}\}+Y(1-k)t^{-1}],\\
H^{10}_1&=&-(te^{-x})^{\frac{1-k}{2}}[2t\{c_{-}b_{-}t^{b_{-}-1}e^{-xb_{-}}
+m^2bt^{b-1}{e^x}^b\nonumber\\
&+&c_{+}b_{+}t^{b_{+}-1}e^{-xb_{+}}\}+(1-k)Y].
\end{eqnarray}
Substituting these values in Eq.(3), we obtain the components of
energy, momentum and energy current densities as
\begin{eqnarray}
\Theta^0_0&=&-\frac{1}{16\pi}[(1-k)(te^{-x})^{\frac{1-k}{2}}
\{c_{-}b_{-}t^{b_{-}-1}e^{-xb_{-}}
-m^2bt^{b-1}e^{xb}\nonumber\\&+&c_{+}b_{+}t^{b_{+}-1}e^{-xb_{+}}\}
+2(te^{-x})^{\frac{1-k}{2}}\{c_{-}b^2_{-}t^{b_{-}-1}e^{-xb_{-}}
+m^2b^2t^{b-1}e^{xb}\nonumber\\&+&c_{+}b^2_{+}t^{b_{+}-1}e^{-xb_{+}}\}
+\frac{(1-k)^2}{2}t^{-\frac{1+k}{2}}e^{-x\frac{(1-k)}{2}}
\{c_{-}(te^{-x})^{b_{-}}\nonumber\\&+&m^2(te^x)^b+c_{+}(te^{-x})^{b_{+}}\}
+(1-k)t^{-\frac{(1+k)}{2}}e^{-x\frac{(1-k)}{2}}\nonumber\\&&
\{c_{-}b_{-}(te^{-x})^{b_{-}}
-m^2b(te^x)^b+c_{+}b_{+}(te^{-x})^{b_{+}}\}].
\end{eqnarray}
\begin{eqnarray}
\Theta^0_1&=&\frac{(te^{-x})^{\frac{1-k}{2}}}{16\pi}[\frac{(1-k)^2}{2}
\{c_{-}(te^{-x})^{b_{-}}+m^2(te^x)^b+c_{+}(te^{-x})^{b_{+}}\}\nonumber\\
&+&(1-k)\{c_{-}b_{-}(te^{-x})^{b_{-}}-m^2b(te^x)^b
+c_{+}b_{+}(te^{-x})^{b_{+}}\}\nonumber\\
&-&t(1-k)\{c_{-}b_{-}t^{b_{-}-1}e^{-xb_{-}}
+m^2bt^{b-1}e^{xb}+c_{+}b_{+}t^{b_{+}-1}e^{-xb_{+}}\}\nonumber\\
&-&2t\{c_{-}b^2_{-}t^{b_{-}-1}e^{-xb_{-}}
-m^2b^2t^{b-1}e^{xb}+c_{+}b^2_{+}t^{b_{+}-1}e^{-xb_{+}}\}],\\
\Theta^1_0&=&\frac{1}{16\pi}[(1-k)^2t^{-\frac{(3+k)}{2}}e^{-x\frac{(1-k)}{2}}
\{c_{-}(te^{-x})^{b_{-}}+m^2(te^x)^b\nonumber\\&+&c_{+}(te^{-x})^{b_{+}}\}
+(1-k)t^{-\frac{(3+k)}{2}}e^{-x\frac{(1-k)}{2}}
\{c_{-}b_{-}(te^{-x})^{b_{-}}-m^2b(te^x)^b
\nonumber\\&+&c_{+}b_{+}(te^{-x})^{b_{+}}\}
-(1-k)t^{-\frac{(1+k)}{2}}e^{-x\frac{(1+k)}{2}}\{c_{-}b_{-}t^{b_{-}-1}e^{-xb_{-}}
\nonumber\\&+&m^2bt^{b-1}e^{xb}+c_{+}b_{+}t^{b_{+}-1}e^{-xb_{+}}\}
-2t^{-\frac{(1+k)}{2}}e^{-x\frac{(1+k)}{2}}\nonumber\\&&
\{c_{-}b^2_{-}t^{b_{-}-1}e^{-xb_{-}}
-m^2b^2t^{b-1}e^{xb}+c_{+}b^2_{+}t^{b_{+}-1}e^{-xb_{+}}\}],\\
\Theta^0_2&=&\Theta^2_0=0=\Theta^0_3=\Theta^3_0.
\end{eqnarray}

In the Landau-Lifshitz prescription, we use the component of
$\ell^{acbd}$
\begin{eqnarray}
\ell^{1001}&=&-\ell^{0011}=(te^{-x})^{1-k}[c_{-}(te^{-x})^{b_{-}}\nonumber\\
&+&m^2(te^x)^b+c_{+}(te^{-x})^{b_{+}}].
\end{eqnarray}
The components of energy and momentum (energy current) densities are
\begin{eqnarray}
L^{00}&=&\frac{1}{16\pi}[(1-k)^2(te^{-x})\{c_{-}(te^{-x})^{b_{-}}
+m^2(te^x)^b+c_{+}(te^{-x})^{b_{+}}\}^2\nonumber\\&+&4(1-k)(te^{-x})^{1-k}
\{c_{-}(te^{-x})^{b_{-}}+m^2(te^x)^b+c_{+}(te^{-x})^{b_{+}}\}\nonumber\\
&\times&\{c_{-}b_{-}(te^{-x})^{b_{-}}
-m^2b(te^x)^b+c_{+}b_{+}(te^{-x})^{b_{+}}\}\nonumber\\
&-&2(te^{-x})^{1-k}\{c_{-}b_{-}(te^{-x})^{b_{-}}
-m^2b(te^x)^b+c_{+}b_{+}(te^{-x})^{b_{+}}\}^2\nonumber\\
&+&2(te^{-x})^{1-k}\{c_{-}b_{-}(te^{-x})^{b_{-}}
-m^2b(te^x)^b+c_{+}b_{+}(te^{-x})^{b_{+}}\}\nonumber\\
&\times&\{c_{-}(te^{-x})^{b_{-}}
+m^2(te^x)^b+c_{+}(te^{-x})^{b_{+}}\}].
\end{eqnarray}
\begin{eqnarray}
\L^{10}&=&L^{01}=\frac{1}{16\pi}[2\{c_{-}(b_{-}+1-k)(te^{-x})^{b_{-}+1-k}
-m^2b(te^x)^b(te^{-x})^{1-k}\nonumber\\&+&m^2(1-k)(te^x)^b(te^{-x})^{1-k}
+c_{+}b_{+}+1-k)(te^{-x})^{b_{+}+1-k}\}\nonumber\\
&\times&\{c_{-}b^2_{-}t^{b_{-}-1}e^{-xb_{-}}-m^2b^2t^{b-1}e^{xb}
+c_{+}b^2_{-}t^{b_{+}-1}e^{-xb_{+}}\}\nonumber\\
&-&(1-k)\{c_{-}(b_{-}+1-k)t^{b_{-}-k}e^{-x(b_{-}+1-k)}\nonumber\\
&+&m^2t^{b-k}e^{-x(1-k-b)}(1-k-b)\nonumber\\
&+&c_{+}(b_{+}+1-k)t^{b_{+}-k}e^{-x(b_{+}+1-k)}\}],\\
L^{20}&=&L^{02}=0=L^{30}=L^{03}.
\end{eqnarray}

For Papapetrou complex, the following components of $N^{abcd}$ are
used
\begin{eqnarray}
N^{0011}&=&-(t^2+1)(t^{-\frac{(1+k)}{2}}e^{-x\frac{(1-k)}{2}})
\nonumber\\&&[c_{-}(te^{-x})^{b_{-}}
+m^2(te^x)^b+c_{+}(te^{-x})^{b_{+}}],\\
N^{0101}&=&e^{-x{\frac{(1-k)}{2}}}(t^{-(1+k)}+t^{-\frac{(1+k)}{2}})
\nonumber\\&&[c_{-}(te^{-x})^{b_{-}}+m^2(te^x)^b+c_{+}(te^{-x})^{b_{+}}].
\end{eqnarray}
Consequently, the components of energy and momentum (energy current)
densities turn out to be
\begin{eqnarray}
\Omega^{00}&=&-\frac{(t^2+1)(t^{-\frac{(1+k)}{2}}
e^{-x\frac{(1-k)}{2}})}{16\pi}[\frac{(1-k)^2}{2}\{c_{-}(te^{-x})^{b_{-}}\nonumber\\
&+&m^2(te^x)^b+c_{+}(te^{-x})^{b_{+}}\}+(1-k)\{c_{-}b_{-}(te^{-x})^{b_{-}}\nonumber\\
&-&m^2b(te^x)^b+c_{+}b_{+}(te^{-x})^{b_{+}}\}+\{c_{-}b^2_{-}(te^{-x})^{b_{-}}\nonumber\\
&-&m^2b^2(te^x)^b+c_{+}b^2_{+}(te^{-x})^{b_{+}}\}].
\end{eqnarray}
\begin{eqnarray}
\Omega^{10}&=&\Omega^{01}=\frac{1}{16\pi}[\frac{(1-k)}{2}e^{-x\frac{(1-k)}{2}}
\{(t^{-(2+k)}(1+k)\nonumber\\&+&t^{-\frac{(3+k)}{2}}\frac{(1+k)}{2})\times
(c_{-}(te^{-x})^{b_{-}}+m^2(te^x)^b\nonumber\\
&+&c_{+}(te^{-x})^{b_{+}})+(t^{-(1+k)}+t^{-\frac{(1+k)}{2}})
\times(c_{-}b_{-}t^{b_{-}-1}e^{-xb_{-}}\nonumber\\&+&m^2bt^{b-1}e^{bx}
+c_{+}b_{+}t^{b_{+}-1}e^{-xb_{+}})\}\nonumber\\&+&e^{-x\frac{(1-k)}{2}}
\{(t^{-(2+k)}(1+k)+t^{-\frac{(3+k)}{2}}\frac{(1+k)}{2})\nonumber\\
&\times&(c_{-}b_{-}t^{b_{-}}e^{-xb_{-}}-m^2bt^be^{bx}
+c_{+}b_{+}t^{b_{+}}e^{-xb_{+}})\nonumber\\&-&(t^{-(1+k)}+t^{-\frac{(1+k)}{2}})
\times(c_{-}b^2_{-}t^{b_{-}-1}e^{-xb_{-}}\nonumber\\&-&m^2b^2t^{b-1}e^{bx}
+c_{+}b^2_{+}t^{b_{+}-1}e^{-xb_{+}})\}],\\
\Omega^{20}&=&\Omega^{02}=0=\Omega^{30}=\Omega^{03}.
\end{eqnarray}

The required component of $K^{bc}_a$ in M\"{o}ller prescription is
\begin{equation}
K^{01}_1=2(te^{-x})^{\frac{(1-k)}{2}}[c_{-}(te^{-x})^{b_{-}}
+m^2(te^x)^b+c_{+}(te^{-x})^{b_{+}}]
\end{equation}
and the components of energy, momentum and energy current densities
become
\begin{eqnarray}
M^0_0&=&0=M^1_0=0,\\
M^0_1&=&-\frac{1}{4\pi}[c_{-}(\frac{1-k}{2}
+b_{-})(te^{-x})^{\frac{(1-k)}{2}+b_{-}}\nonumber\\
&+&\frac{(1-k)}{2}m^2(te^x)^b (te^{-x})^{\frac{(1-k)}{2}+b_{+}}
-m^2b(te^x)^b(te^{-x})^{\frac{(1-k)}{2}}\nonumber\\
&+&c_{+}(\frac{1-k}{2}+b_{+})(te^{-x})^{\frac{(1-k)}{2}+b_{+}}],\\
M^0_2&=&0=M^2_0=M^0_3=M^3_0.
\end{eqnarray}
This gives constant energy and momentum.

\section{Wainwright-Marshman Solution}

The line element of Wainwright-Marshman solution [50] is given as
follows
\begin{equation}
ds^2=t^{2m}e^n(dt^2-dx^2)-t^{\frac{1}{2}}dy^2-2\omega
t^{\frac{1}{2}}dydz -(t^{\frac{1}{2}}\omega^2+t^{\frac{3}{2}})dz^2,
\end{equation}
where $\omega=\omega(t-x)$ is an arbitrary function, $n=n(t-x)$ is
determined according to $n'=(\omega')^2$ and $m$ is constant. For
this metric, energy-momentum turns out to be constant in the
Einstein, Landau-Lifshitz and Papapetrou prescriptions. In the
M\"{o}ller's prescription, the components of energy, momentum and
energy current densities are
\begin{eqnarray}
M^0_0&=&\frac{tn''}{8\pi},\\
M^1_0&=&-\frac{n'}{8\pi},\\
M^0_1&=&0=M^0_2=M^2_0=M^0_3=M^3_0.
\end{eqnarray}
This shows that energy and momentum can be constant only if $n$ is
constant.

\section{Summary and Discussion}

This paper continues the investigation of comparing various
distributions presented in the literature. It is devoted to discuss
the burning problem of energy-momentum in the framework of GR and
four different energy-momentum complexes have been used to find the
energy-momentum distribution. We have applied the prescriptions of
Einstein, Landau-Lifshitz, Papapetrou and M\"{o}ller to investigate
energy-momentum distribution for various spacetimes. The summary of
the results (only non-zero quantities) can be given in the form of
tables in the following:

\vspace{0.2in}
\begin{center}
\textbf{Table 1(a) Gravitational Waves:  Einstein Complex}
\vspace{0.2in}

\begin{tabular}{|c|c|}

\hline{\bf Energy-Momentum Densities}&{\bf Expressions}\\
\hline
$\Theta^0_0$ &$\frac{e^{-(2M+3U)}}{8 \pi}(U''-2U'M'-3U'^2)$\\
\hline $\Theta^1_0$&$\frac{e^{-(2M+3U)}}{8
\pi}[U'(2\dot{M}+3\dot{U})-\dot{U'}]$\\
\hline
$\Theta^0_1$&$\frac{e^{-U}}{8\pi}(\dot{U'}-\dot{U}U')$\\
\hline

\end{tabular}
\end{center}

\begin{center}
\textbf{Table 1(b) Gravitational Waves: Landau-Lifshitz Complex}
\vspace{0.2in}

\begin{tabular}{|c|c|}

\hline{\bf Energy-Momentum Densities}&{\bf Expressions}\\
\hline $L^{00}$& $
\frac{e^{-2U}}{8\pi}(U''-2U'^2)$\\
\hline $L^{10}=L^{01}$ &$
\frac{e^{-2U}}{8\pi}(2U'\dot{U}-\dot{U'})$\\
\hline
\end{tabular}
\end{center}
\vspace{0.2in}
\begin{center}
\textbf{Table 1(c) Gravitational Waves: Papapetrou Complex}
\vspace{0.2in}

\begin{tabular}{|c|c|}

\hline{\bf Energy-Momentum Densities}&{\bf Expressions}\\
\hline $\Omega^{00}$& $
\frac{e^{-U}}{8\pi}(U''-U'^2)$\\
\hline $\Omega^{10}=\Omega^{01}$& $\frac{e^{-U}}{8\pi}(\dot{U}U'-\dot{U'})$\\
\hline
\end{tabular}
\end{center}
\vspace{0.2in}
\begin{center}
\textbf{Table 1(d) Gravitational Waves: M\"{o}ller Complex}
\vspace{0.2in}

\begin{tabular}{|c|c|}

\hline{\bf Energy-Momentum Densities}&{\bf Expressions}\\
\hline $M^0_0$& $
\frac{e^{-U}}{8\pi}(M'U'-M'')$\\
\hline $M^1_0$& $\frac{e^{-U}}{8\pi}(\dot{M'}-\dot{U}M')$\\
\hline
\end{tabular}
\end{center}

\vspace{0.2in}
\begin{center}
\textbf{Table 2(a) Ferrari-Ibanez Degenerate Solution: Einstein
Complex}

\vspace{0.2in}

\begin{tabular}{|c|c|}

\hline{\bf Energy-Momentum Densities}&{\bf Expressions}\\
\hline
$\Theta^0_0$&$\frac{1}{8 \pi}(\cos t\cos z)$\\
\hline
$\Theta^3_0$&$\frac{1}{8\pi}(\sin t\sin z)$\\
\hline
$\Theta^0_3$&$-\frac{1}{8\pi}(\sin t\sin z)$\\
\hline
\end{tabular}
\end{center}
\vspace{0.2in}

\begin{center}
\textbf{Table 2(b) Ferrari-Ibanez Degenerate Solution:
               Landau-Lifshitz Complex}
\vspace{0.2in}

\begin{tabular}{|c|c|}

\hline{\bf Energy-Momentum Densities}&{\bf Expressions}\\
\hline $L^{00}$& $
\frac{1}{8\pi}(\cos t\cos z)$\\
\hline $L^{30}=L^{03}$ &$
\frac{1}{16\pi}(\sin 2 t\sin 2 z)$\\
\hline
\end{tabular}
\end{center}

\vspace{0.2in}
\begin{center}
\textbf{Table 2(c) Ferrari-Ibanez Degenerate Solution: Pappetrou
         Complex}
\vspace{0.2in}

\begin{tabular}{|c|c|}

\hline{\bf Energy-Momentum Densities}&{\bf Expressions}\\
\hline $\Omega^{00}$& $
\frac{1}{8\pi}(\cos t\cos z)$\\
\hline $\Omega^{30}=\Omega^{03}$& $\frac{1}{8\pi}(\sin t\sin z)$\\
\hline
\end{tabular}
\end{center}

\begin{center}
\textbf{Table 2(d) Ferrari-Ibanez Degenerate Solution:
                M\"{o}ller Complex}
\vspace{0.2in}

\begin{tabular}{|c|c|}

\hline{\bf Energy-Momentum Densities}&{\bf Expressions}\\

\hline $M^0_3$& $-\frac{1}{4\pi}\cos t\sin z\frac{(1-\sin
t)^\frac{1}{2}}{(1+\sin t)^\frac{1}{2}}$\\
\hline
\end{tabular}
\end{center}

\vspace{0.2in}
\begin{center}
{\bf {\small Table 3(a)}} {\small \textbf{Senovilla-Vera Dust
Solution: Einstein Complex}}

\vspace{0.2in}
\begin{tabular}{|c|c|}

\hline{\bf E-M Densities}&{\bf Expressions}\\
\hline $\Theta^0_0$ &$
\begin{array}{c}-\frac{1}{16\pi}[(1-k)(te^{-x})^{\frac{1-k}{2}}
\{c_{-}b_{-}t^{b_{-}-1}e^{-xb_{-}} \\
-m^2bt^{b-1}e^{xb}+c_{+}b_{+}t^{b_{+}-1}e^{-xb_{+}}\}
+2(te^{-x})^{\frac{1-k}{2}} \\
\{c_{-}b^2_{-}t^{b_{-}-1}e^{-xb_{-}} +m^2b^2t^{b-1}e^{xb}
\\ +c_{+}b^2_{+}t^{b_{+}-1}e^{-xb_{+}}\}
+\frac{(1-k)^2}{2}t^{-\frac{1+k}{2}}e^{-x\frac{(1-k)}{2}} \\
\{c_{-}(te^{-x})^{b_{-}}+m^2(te^x)^b+c_{+}(te^{-x})^{b_{+}}\}
\\ +(1-k)t^{-\frac{(1+k)}{2}}
e^{-x\frac{(1-k)}{2}}\{c_{-}b_{-}(te^{-x})^{b_{-}}\\ -
m^2b(te^x)^b+c_{+}b_{+}(te^{-x})^{b_{+}}\}]\end{array}$\\
\hline $\Theta^1_0$& $\begin{array}{c}
\frac{1}{16\pi}[(1-k)^2t^{-\frac{(3+k)}{2}}e^{-x\frac{(1-k)}{2}}
\{c_{-}(te^{-x})^{b_{-}}\\+m^2(te^x)^b+c_{+}(te^{-x})^{b_{+}}\}
+(1-k)t^{-\frac{(3+k)}{2}}e^{-x\frac{(1-k)}{2}} \\
\{c_{-}b_{-}(te^{-x})^{b_{-}}-m^2b(te^x)^b
+c_{+}b_{+}(te^{-x})^{b_{+}}\}\\
-(1-k)t^{-\frac{(1+k)}{2}}e^{-x\frac{(1+k)}{2}}\{c_{-}b_{-}t^{b_{-}-1}e^{-xb_{-}}
\\ +m^2bt^{b-1}e^{xb}+c_{+}b_{+}t^{b_{+}-1}e^{-xb_{+}}\}\\
-2t^{-\frac{(1+k)}{2}}e^{-x\frac{(1+k)}{2}}\{c_{-}b^2_{-}t^{b_{-}-1}e^{-xb_{-}}
-m^2b^2t^{b-1}e^{xb}\\
+c_{+}b^2_{+}t^{b_{+}-1}e^{-xb_{+}}\}]\end{array}$\\
\hline $\Theta^0_1$& $\begin{array}{c}
\frac{(te^{-x})^{\frac{1-k}{2}}}{16\pi}[\frac{(1-k)^2}{2}
\{c_{-}(te^{-x})^{b_{-}}+m^2(te^x)^b\\+c_{+}(te^{-x})^{b_{+}}\}+
(1-k)\{c_{-}b_{-}(te^{-x})^{b_{-}}\\-m^2b(te^x)^b
+c_{+}b_{+}(te^{-x})^{b_{+}}\}\\
-t(1-k)\{c_{-}b_{-}t^{b_{-}-1}e^{-xb_{-}}
+m^2bt^{b-1}e^{xb}\\+c_{+}b_{+}t^{b_{+}-1}e^{-xb_{+}}\}
-2t\{c_{-}b^2_{-}t^{b_{-}-1}e^{-xb_{-}}\\
-m^2b^2t^{b-1}e^{xb}+c_{+}b^2_{+}t^{b_{+}-1}e^{-xb_{+}}\}]\end{array}$\\
\hline
\end{tabular}
\end{center}
\newpage
\begin{center}
{\bf {\small Table 3(b)}} {\small \textbf{Senovilla-Vera Dust
Solution: Landau-Lifsihtz Complex}}

\vspace{0.2in}
\begin{tabular}{|c|c|}

\hline{\bf E-M Densities}&{\bf Expressions}\\
\hline $L^{00}$& $\begin{array}{c}
\frac{1}{16\pi}[(1-k)^2(te^{-x})\{c_{-}(te^{-x})^{b_{-}}
+m^2(te^x)^b\\+c_{+}(te^{-x})^{b_{+}}\}^2+4(1-k)(te^{-x})^{1-k}
\{c_{-}(te^{-x})^{b_{-}}\\+m^2(te^x)^b+c_{+}(te^{-x})^{b_{+}}\}
\{c_{-}b_{-}(te^{-x})^{b_{-}}\\
-m^2b(te^x)^b+c_{+}b_{+}(te^{-x})^{b_{+}}\} -2(te^{-x})^{1-k}\\
\{c_{-}b_{-}(te^{-x})^{b_{-}}
-m^2b(te^x)^b+c_{+}b_{+}(te^{-x})^{b_{+}}\}^2
\\+2(te^{-x})^{1-k}\{c_{-}b_{-}(te^{-x})^{b_{-}}
-m^2b(te^x)^b\\+c_{+}b_{+}(te^{-x})^{b_{+}}\}
\{c_{-}(te^{-x})^{b_{-}}
+m^2(te^x)^b\\+c_{+}(te^{-x})^{b_{+}}\}]\end{array}$\\
\hline $L^{10}=L^{01}$ &$\begin{array}{c}
\frac{1}{16\pi}[2\{c_{-}(b_{-}+1-k)(te^{-x})^{b_{-}+1-k}
\\-m^2b(te^x)^b(te^{-x})^{1-k}+m^2(1-k)(te^x)^b(te^{-x})^{1-k}
\\+c_{+}(b_{+}+1-k)(te^{-x})^{b_{+}+1-k}\}
\{c_{-}b^2_{-}t^{b_{-}-1}e^{-xb_{-}}\\-m^2b^2t^{b-1}e^{xb}
+c_{+}b^2_{-}t^{b_{+}-1}e^{-xb_{+}}\}\\
-(1-k)\{c_{-}(b_{-}+1-k)t^{b_{-}-k}e^{-x(b_{-}+1-k)}
\\+m^2t^{b-k}e^{-x(1-k-b)}(1-k-b)
\\+c_{+}(b_{+}+1-k)t^{b_{+}-k}e^{-x(b_{+}+1-k)}\}]\end{array}$\\
\hline
\end{tabular}
\end{center}

\vspace{0.2in}

\begin{center}
{\bf {\small Table 3(c)}} {\small \textbf{Senovilla-Vera Dust
Solution: Papapetrou Complex}}

\vspace{0.2in}
\begin{tabular}{|c|c|}

\hline{\bf E-M Densities}&{\bf Expressions}\\
\hline $\Omega^{00}$& $\begin{array}{c}
-\frac{(t^2+1)(t^{-\frac{(1+k)}{2}}
e^{-x\frac{(1-k)}{2}})}{16\pi}[\frac{(1-k)^2}{2}\{c_{-}(te^{-x})^{b_{-}}
\\+m^2(te^x)^b+c_{+}(te^{-x})^{b_{+}}\}+(1-k)\{c_{-}b_{-}(te^{-x})^{b_{-}}
\\-m^2b(te^x)^b+c_{+}b_{+}(te^{-x})^{b_{+}}\}+\{c_{-}b^2_{-}(te^{-x})^{b_{-}}
\\-m^2b^2(te^x)^b+c_{+}b^2_{+}(te^{-x})^{b_{+}}\}]\end{array}$\\
\hline $\Omega^{10}=\Omega^{01}$& $\begin{array}{c}
\frac{1}{16\pi}[\frac{(1-k)}{2}e^{-x\frac{(1-k)}{2}}
\{(t^{-(2+k)}(1+k)\\+t^{-\frac{(3+k)}{2}}\frac{(1+k)}{2})
(c_{-}(te^{-x})^{b_{-}}+m^2(te^x)^b\\
+c_{+}(te^{-x})^{b_{+}})+(t^{-(1+k)}+t^{-\frac{(1+k)}{2}})
(c_{-}b_{-}t^{b_{-}-1}e^{-xb_{-}}\\+m^2bt^{b-1}e^{bx}
+c_{+}b_{+}t^{b_{+}-1}e^{-xb_{+}})\}\\+e^{-x\frac{(1-k)}{2}}
\{(t^{-(2+k)}(1+k)+\frac{(1+k)}{2}t^{-\frac{(3+k)}{2}})\\
(c_{-}b_{-}t^{b_{-}}e^{-xb_{-}}-m^2bt^be^{bx}
+c_{+}b_{+}t^{b_{+}}e^{-xb_{+}})\\-(t^{-(1+k)}+t^{-\frac{(1+k)}{2}})
(c_{-}b^2_{-}t^{b_{-}-1}e^{-xb_{-}}\\-m^2b^2t^{b-1}e^{bx}
+c_{+}b^2_{+}t^{b_{+}-1}e^{-xb_{+}})\}]\end{array}$\\
\hline
\end{tabular}
\end{center}
\vspace{0.2in}
\begin{center}
{\bf {\small Table 3(d)}} {\small \textbf{Senovilla-Vera Dust
Solution: M\"{o}ller Complex}}

\vspace{0.2in}
\begin{tabular}{|c|c|}

\hline{\bf E-M Densities}&{\bf Expressions}\\
\hline $M^0_1$& $\begin{array}{c}
-\frac{1}{4\pi}[c_{-}(\frac{1-k}{2}
+b_{-})(te^{-x})^{\frac{(1-k)}{2}+b_{-}}\\+\frac{(1-k)}{2}m^2(te^x)^b
(te^{-x})^{\frac{(1-k)}{2}+b_{+}}
\\-m^2b(te^x)^b(te^{-x})^{\frac{(1-k)}{2}}
+c_{+}(\frac{1-k}{2}+b_{+})(te^{-x})^{\frac{(1-k)}{2}+b_{+}}]\end{array}$\\
\hline
\end{tabular}
\end{center}
From the above tables, it is concluded that the energy-momentum
densities turn out to be finite and well-defined in all the
prescriptions for the spacetimes under consideration. We find that
the three prescriptions, i.e., Einstein, Landau-Lifshitz and
Papapetrou complexes provide the same energy distribution for the
special class of Ferrari-Ibanez degenerate solution while M\"{o}ller
prescription gives constant energy and momentum (Table 2). If we
take $t$ or $z=\pi/2$, energy becomes constant and for $t=0=z$,
momentum turns out to be constant in the remaining prescriptions as
well. For the Wainwright-Marshman solution, energy and momentum is
constant in all the prescriptions except M\"{o}ller where energy and
momentum densities are constant for $n$ to be constant. The metric
exhibiting asymptotic silence-breaking singularities, Senovilla-Vera
dust solution yields constant energy-momentum only in M\"{o}ller
prescription while it has different non-vanishing energy-momentum
densities in the remaining prescriptions. The energy-momentum
densities turn out to be different for gravitational waves in all
the prescriptions (Table 1).

It is worth mentioning that the results of energy-momentum
distribution for the two examples turn out to be same in all the
prescriptions. These results justify the viewpoint of Virbhadra and
his collaborators [26-37] that different energy-momentum complexes
may provide some basis to define a unique quantity. However, the
remaining two examples give different energy-momentum densities in
different prescriptions. This difference is due to the fact that
different energy-momentum complexes, which are pseudo-tensors, are
not covariant objects. This is in accordance with the equivalence
principle [2] which implies that the gravitational field cannot be
detected at a point. This also supports the viewpoint of Cooperstock
[3] that energy can not be localized. Notice that each expression
may have a geometrically and physically clear significance
associated with the boundary conditions.

\newpage
{\bf \large References}

\begin{description}

\item{[1]} M\"{o}ller, C.: {\it The Theory of Relativity} (Oxford
University Press, 1957).

\item{[2]} Misner, C.W., Thorne, K.S. and Wheeler, J.A.: {\it
Gravitation} (W. H. Freeman and Co., NY 1973)603.

\item{[3]} Cooperstock, F.I. and Sarracino, R.S.: J. Phys. A: Math. Gen.
\textbf{11}(1978)877.

\item{[4]} Trautman, A.: \textit{Gravitation: An Introduction to Current Research,}
ed. Written, L. (Wiley, New York, 1962).

\item{[5]} Tolman, R.C.: \textit{Relativity, Thermodynamics and Cosmology}
(Oxford University Press, London, 1934).

\item{[6]} Landau, L.D. and Lifshitz, E.M.: \textit{The Classical Theory of
Fields} (Pergamon, Oxford, 1980).

\item{[7]} Papapetrou, A.: \textit{Proc. R. Irish Acad.} \textbf{A52}(1948)11.

\item{[8]} Bergmann, P. G. and Thompson, R.: Phys. Rev.
\textbf{89}(1953)400.

\item{[9]} Weinberg, S.: \textit{Gravitation and Cosmology} (Wiley, New York, 1972).

\item{[10]} Bondi, H.: \textit{Proc. R. Soc. London}
\textbf{A427}(1990)249.

\item{[11]} Penrose, R.: \textit{Proc. R. Soc. London}
\textbf{A388}(1982)457; GR 10 \textit{Conference} eds. Bertotti, B.
de Felice, F. and Pascolini, A. Padova, \textbf{1}(1983)607.

\item{[12]} Brown, J.D. and York, Jr. J.W.: Phys. Rev. \textbf{D47}(1993)1407.

\item{[13]} Hayward, S.A.: Phys. Rev. \textbf{D497}(1994)831.

\item{[14]} Berqvist, G.: Class. Quantum Grav. \textbf{9}(1992)1753.

\item{[15]} Chang, C.C., Nester, J.M. and Chen, C.: Phys. Rev.
Lett. \textbf{83}(1999)1897.

\item{[16]} M\"{o}ller, C.: Ann. Phys. (NY) {\bf 4}(1958)347.

\item{[17]} M\"{o}ller, C.: Ann. Phys. (NY) \textbf{12}(1961)118.

\item{[18]} Lessner, G.: Gen. Rel. Grav. \textbf{28}(1996)527.

\item{[19]} Kovacs, D.: Gen. Rel. Grav. \textbf{17}(1985)927.

\item{[20]} Novotny, J.: Gen. Rel. Grav. \textbf{19}(1987)1043.

\item{[21]} Komar, A.: Phys. Rev. \textbf{113}(1959)934.

\item{[22]} Scheidegger, A. E.: Rev. Mod. Phys. \textbf{25}(1953)451.

\item{[23]} Rosen, N.: Helv. Phys. Acta. Suppl. \textbf{4}(1956)171.

\item{[24]} Rosen, N.: Phys. Rev. \textbf{110}(1958)291.

\item{[25]} Rosen, N. and Virbhadra, K.S: Gen. Rel. Grav.
\textbf{26}(1993)429.

\item{[26]}  Virbhadra, K.S.: Phys. Rev. {\bf D41}(1990)1086.

\item{[27]}  Virbhadra, K.S.: {\bf D42}(1990)1066.

\item{[28]} Virbhadra, K.S.: Phys. Rev. {\bf D60}(1999)104041.

\item{[29]} Virbhadra, K.S.: Phys. Rev. {\bf D42}(1990)2919.

\item{[30]} Virbhadra, K.S. and Parikh, J.C.: Phys. Lett. {\bf B317}(1993)312.

\item{[31]} Virbhadra, K.S. and Parikh, J.C.: Phys. Lett. {\bf B331}(1994)302.

\item{[32]} Aguirregabiria, J.M., Chamorro, A. and Virbhadra, K.S.:
Gen. Relativ. Gravit. {\bf 28}(1996)1393.

\item{[33]} Xulu, S.S.: Int. J. of Mod. Phys.
\textbf{A15}(2000)1979.

\item{[34]} Xulu, S.S.: Mod. Phys. Lett. \textbf{A15}(2000)1151.

\item{[35]} Xulu, S.S.: Astrophys. Space Sci. \textbf{283}(2003)23.

\item{[36]} Chamorro, A. and Virbhardra, K.S.: Int. J. of Mod. Phys.
\textbf{D5}(1994)251.

\item{[37]} Xulu, S.S.: Int. J. of Mod. Phys. \textbf{D7}(1998)773.

\item{[38]} Sharif, M.: Int. J. of Mod. Phys. {\bf A17}(2002)1175;

\item{[39]} Sharif, M.: Int. J. of Mod. Phys. {\bf A18}(2003)4361.

\item{[40]} Sharif, M.: Int. J. of Mod. Phys. {\bf A19}(2004)1495.

\item{[41]} Sharif, M.: Int. J. of Mod. Phys. {\bf D13}(2004)1019.

\item{[42]} Komar, A.: Phys. Rev. \textbf{127}(1962)1411.

\item{[43]} Komar, A.: Phys. Rev. \textbf{129}(1963)1873.

\item{[44]} Penrose, R.: \textit{Proc. R. Soc. London}
\textbf{A381}(1982)53.

\item{[45]} Penrose, R.: \textit{In Asymptotic Behavior of Mass and
Spacetime Geometry}, ed. Flaherty, J.F. (Springer, Berlin, 1984).

\item{[46]} Feinstein, A. and Griffiths, J.B.: Class. Quantum
Grav. \textbf{11}(1994)L109.

\item{[47]} Alekseev, G.A. and Griffiths, J.B.: Class. Quantum
Grav. \textbf{13}(1996)2191.

\item{[48]} Griffiths, J. B.: \textit{Colliding Plane Wave in General
Relativity} (Oxford University Press, 1991).

\item{[49]} Senovilla, M.M.J. and Vera, R.: Class. Quantum Grav. \textbf{4}(1997)3481.

\item{[50]} Wainwright, J. and Marshman, J.B.: Phys. Lett.
\textbf{A72}(1979)275.

\end{description}

\end{document}